\documentclass{aa}

\usepackage{graphicx}
\usepackage{txfonts}
\begin{document} 
\title{Investigation of Inverse Velocity Dispersion in a Solar Energetic Particle Event Observed by Solar Orbiter}

   \author{Zheyi Ding          \inst{1}
           \and
           Robert F. Wimmer-Schweingruber \inst{1}
           \and
            Alexander Kollhoff \inst{1}
           \and
           Patrick Kühl \inst{1}
           \and 
           Liu Yang \inst{1}           
           \and
           Lars Berger \inst{1}
           \and
            Athanasios Kouloumvakos \inst{2}
           \and
            Nicolas Wijsen \inst{3}
           \and
           Jingnan Guo \inst{4}
           \and
           Daniel Pacheco\inst{4}
           \and
           Yuncong Li \inst{4}
           \and
           Manuela Temmer \inst{5}    
           \and
           Javier Rodriguez-Pacheco \inst{6}
           \and 
           Robert C. Allen \inst{7}  
           \and
           George C. Ho \inst{7}         
           \and
           Glenn M. Mason \inst{2}
           \and 
           Zigong Xu  \inst{8}
           \and
           Sindhuja G. \inst{1}
          }

   \institute{ 
             \inst{1} Institute of Experimental and Applied Physics, Kiel University, Leibnizstrasse 11, D-24118 Kiel, Germany   \email{ding@physik.uni-kiel.de}\\
             \inst{2} Johns Hopkins Applied Physics Lab, Laurel, MD 20723, USA \\             
             \inst{3} Centre for mathematical Plasma Astrophysics, KU Leuven Campus Kulak, 8500 Kortrijk, Belgium\\
             \inst{4} National Key Laboratory of Deep Space Exploration/School of Earth and Space Sciences, University of Science and Technology of China, Hefei 230026, China \\        
             \inst{5} Institute of Physics, University of Graz, Graz, Austria\\
             \inst{6} Universidad de Alcalá, Alcalá de Henares 28805, Spain \\   
             \inst{7} Southwest Research Institute, San Antonio, TX 78228, USA\\
             \inst{8} California Institute of Technology, MC 290-17, Pasadena, CA 91125, USA
             }

   \date{Received ; accepted }

 
\abstract
{Solar Energetic Particle (SEP) events provide crucial insights into particle acceleration and transport mechanisms in the heliosphere. Inverse velocity dispersion (IVD) events, characterized by higher-energy particles arriving later than lower-energy particles, challenge the classical understanding of SEP events and are increasingly observed by spacecraft, such as Parker Solar Probe (PSP) and Solar Orbiter (SolO). However, the mechanisms underlying IVD events remain poorly understood.}
{This study aims to investigate the physical processes responsible for long-duration IVD events by analyzing the SEP event observed by SolO on 2022 June 7.  We explore the role of evolving shock connectivity, particle acceleration at interplanetary (IP) shocks, and cross-field transport in shaping the observed particle profiles.} 
{We utilize data from Energetic Particle Detector (EPD) suite onboard SolO to analyze the characteristics of the IVD, and model the event using the Heliospheric Energetic Particle Acceleration and Transport (HEPAT) model. The simulations track evolutions of shock properties, particle acceleration and transport to assess the influence of shock expansion, shock connectivity, and transport processes on the formation of IVD.
}
{The IVD event exhibited a distinct and long-duration IVD signature, across proton energies from 1 to 20 MeV and lasting for approximately 10 hours. Heavy ions exhibited varying nose energies, defined as the energy corresponding to the first-arriving particles. Simulations suggest that evolving shock connectivity and the evolution of shock play a primary role in the IVD signature, with SolO transitioning from shock flank to nose over time, resulting in a gradual increase in maximum particle energy along the field line. Furthermore,  model results show that limited cross-field diffusion can influence both the nose energy and the duration of the IVD event.  
}
{This study demonstrates that long-duration IVD events are primarily driven by evolving magnetic connectivity along a non-uniform shock that evolves over time, where the connection moves to more efficient acceleration sites as the shock propagates farther from the Sun.  Other mechanisms, such as acceleration time at the shock, may also contribute to the observed IVD features. The interplay of these factors remains an open question, warranting further investigation on other events.  }

   \titlerunning {Investigating the IVD event observed by SolO}
   \authorrunning{Ding et al.}
   \maketitle
%

\section{Introduction}
\label{sec:intro}

Solar Energetic Particle (SEP) events, associated with solar eruptions such as flares and coronal mass ejections (CMEs),  provide critical insights into particle acceleration and transport mechanisms. Large SEP events are typically associated with shock waves driven by CME that can accelerate particles to relativistic energies \citep[see the reviews by][]{Desai2016,Reames1999}. These shocks are efficient sites for particle acceleration, primarily considered through the mechanism of diffusive shock acceleration \citep[DSA;][]{Axford1977, Drury+1983}. In the DSA mechanism, particle transport around the shock is governed by diffusion, with particles freely traversing the shock front. This diffusion arises from the scattering of particles by magnetic field irregularities convected by the ambient plasma flow. Every time particles cross the shock, they receive a net energy gain.  This iterative process accelerates particles to higher energies, thus the maximum particle energy is controlled by the scattering efficiency and the finite acceleration time.  After accelerated particles escape from the traveling shock front, they propagate along the interplanetary magnetic field (IMF).  During their propagation,  particles undergo a series of transport processes, including pitch-angle scattering, adiabatic momentum changes, and magnetic focusing/mirroring, which shape their spatio-temporal and energy distributions \citep{Li+2003,Wijsen2019,Hu+etal+2018}. In addition to field-aligned propagation, particles may diffuse perpendicular to the magnetic field, enabling cross-field transport and allowing particles to spread widely in longitude and latitude \citep[e.g., ][]{dwyer1997perpendicular,strauss2017,wang2012effects}. Such cross-field transport is essential for explaining the widespread SEP events, especially at locations magnetically disconnected from the initial shock region \citep[e.g., ][]{Li2021,ding2022A&A...668A..71D}. Ultimately, the combination of particle acceleration at the traveling shock and particle propagation along and across field lines governs the distribution of SEP intensity throughout the heliosphere, influencing the time-intensity profiles recorded by spacecraft at different vantage points.

If particles are released from the shock simultaneously, high-energy particles arrive at the observer earlier than lower-energy particles. This behavior, known as velocity dispersion (VD), is often observed in the initial phase of SEP events \citep[see e.g., ][]{McCracken1970SSRv,Reames1997ApJ,Kollhoff2021A&A,wimmer2023unusually}. Velocity dispersion analysis (VDA) assumes the scatter-free propagation of particles along magnetic field lines, allowing their energy-dependent arrival times to be used for estimating the particle release time and the path length \citep{Tylka2003, Laitinen2015}.  Recently, a rare event characterized by a prominent feature of inverse velocity dispersion (IVD) was observed by Parker Solar Probe \citep[PSP;][]{Fox2016} at a solar distance of 15 solar radii \citep{Cohen2024}, where higher-energy particles arrived later than lower-energy particles.  In the dynamic spectrum, this event exhibits a well-defined nose structure, with the nose energy for protons around $1\;\mathrm{MeV}$. The nose energy corresponds to the energy channel of the first-arriving particles. In that event, the duration of the IVD is approximately half an hour, measured from the onset time of the nose energy to the onset of the highest-energy particles. \cite{Cohen2024} suggest that this phenomenon can be well explained by the DSA process. It requires time for particles to be accelerated to high energies. Therefore, if an observer is sufficiently close to a shock undergoing particle acceleration, lower-energy particles, which have undergone a shorter acceleration process, are likely to reach the observer first, followed by higher-energy particles.
Additionally, \cite{Kouloumvakos2025} suggest that the observed IVD is attributed to a relatively slow, ongoing particle acceleration process occurring at the flank of the expanding shock wave intercepted by PSP. This slow acceleration process is due to  PSP's initial magnetic connection to a weak region of the shock, which gradually strengthens as the shock expands. 
It is reasonable that the continuous particle acceleration and the initial expansion of the shock collectively contribute to the observed IVD. Given that PSP is located so close to the Sun, the short duration of the IVD minimizes the influence of magnetic connectivity variations and particle transport effects. This allows for a unique observational perspective, where PSP can effectively probe time-dependent particle acceleration and the evolving shock strength at a nearly fixed location along the shock front. Such observations provide critical insights into the early-stage shock dynamics and the microphysics of SEP acceleration near the Sun.

{Extending from the scenario suggested above, we note that particle acceleration at shocks is a dynamic process resulting from the time‐dependent interplay of multiple factors. As shown by \cite{Zank+etal+2000}, the acceleration timescale at a shock depends not only on the varying shock strength but also on the concurrent weakening of the magnetic field with heliocentric distance ($R$) which typically decreases as $R^{-2}$ in the young solar wind. This establishes a direct competition: the rate at which the shock strength evolves must be balanced against the decay of the magnetic field to determine the efficiency of particle acceleration. Furthermore, shock geometry and the associated intensity of upstream waves further complicate the determination of the acceleration time.  A key factor for particle acceleration at quasi-parallel shocks is the excitation of the upstream scattering wave by the energetic particles escaping upstream from the shock. For quasi-perpendicular shocks, wave excitation is quenched, and particle scattering must therefore rely on the ambient solar wind turbulence that is convected into the shock. As suggested in \cite{Zank2006}, the inclusion of wave excitation at quasi-parallel shocks can reduce their timescale even further, resulting in faster acceleration, though quasi‐perpendicular shocks yield significantly shorter acceleration timescales than quasi‐parallel shocks. In addition, injection energy plays an important role in shock acceleration: quasi-perpendicular shocks require a higher injection energy compared to quasi-parallel shocks \citep{Zank2006,Li+2012,Ding2023JGRA..12831502D}. These physical processes have been widely examined using the PATH model \citep{Zank+etal+2000,Li+2003,Li+2005ions,Rice+2003} and two-dimensional iPATH model \citep{Hu+etal+2017},  providing comprehensive insights into the particle acceleration and transport in gradual SEP events.
 }

Interestingly, IVD events are not exclusively observed by PSP close to the Sun. As solar activity increases with the rise toward the solar maximum of cycle 25,  Solar Orbiter (SolO; \citealt{Muller2020}) has detected more than 10 so-called IVD events at varying radial distances over the past few years (see \cite{Li+2025} for details). These events exhibit distinct nose energies, ranging from a few $\mathrm{MeV}$ to tens of $\mathrm{MeV}$, and display varying IVD durations. In comparison to the PSP event on 2022 September 5, where the IVD lasted approximately only half an hour, some SolO events display the IVD duration beyond 10 hours. This extended duration provides important insights that might challenge interpretations of IVD solely as a result of acceleration time at the shock front.  In-situ observations of SEP events inherently reflect the interplay between particle acceleration and transport processes.  As the shock propagates outward, the magnetic connectivity between the observer and the shock naturally evolves, suggesting that long-duration IVD events may not be explained by acceleration processes alone.  In such cases, shock connectivity, the evolution of the shock itself, and transport effects, such as cross-field diffusion, cannot be simply disregarded.

The long-duration IVD events observed by SolO highlight the need for more detailed study to fully understand the origins and variations of IVD events. In this work, we focus on analyzing a well-defined IVD event observed by SolO on 2022 June 7,  which exhibits a distinct nose structure and a long IVD duration, providing an ideal example to investigate the underlying mechanisms responsible for generating IVD signatures. A more comprehensive list of IVD events can be found in \cite{Li+2025}.  
Section~\ref{sec:methods} provides an overview of the measurements used in this study, and the modelling framework employed to simulate this event. Section~\ref{sec:results} presents the analysis of both observational data and model results, offering qualitative insights on the formation of IVD events. The main conclusions are summarized in Sect.~\ref{sec:conclu}.

\section{Data and Methods}\label{sec:methods}

This study utilizes data collected by instruments onboard the Solar Orbiter. Specifically, SEP measurements are obtained using the Energetic Particle Detector suite (EPD; \citealt{Pacheco2020A&A...642A...7R, Wimmer-Schweingruber2021A&A...656A..22W}), which includes the Supra-Thermal Electron and Proton sensor (STEP), the Electron Proton Telescope (EPT), the High Energy Telescope (HET) and the Suprathermal Ion Spectrograph (SIS).  
 These instruments provide data for electron energies ranging from a few keV to tens MeV, and for ions, from a few keV nucleon$^{-1}$ up to over $100\;$MeV nucleon$^{-1}$. EPD provides coverage in four viewing directions through EPT and HET, one through STEP, and two through SIS. A detailed description of the EPD instrument, including the energy ranges for each sensor, particle species, and fields of view, is available in \citet{Pacheco2020A&A...642A...7R}.  The averaged particle intensity from different telescopes is used in this study in order to compare with the omni-directional intensity in the model. It is important to note that STEP and EPT do not explicitly distinguish between various ion species. The measured fluxes are typically interpreted as being dominated by protons, which are the most abundant ion species in large SEP events. Magnetic field data are obtained through measurements from the SolO Magnetometer (MAG; \citealt{Horbury2020A&A...642A...9H}), and plasma data are collected using the Proton-Alpha Sensor in Solar Wind Analyzer (SWA-PAS; \citealt{Owen2020A&A...642A..16O}). 

To simulate the observed IVD features, we employ the three-dimensional (3D) SEP model, named Heliospheric Energetic Particle Acceleration and Transport \citep[HEPAT;][]{Ding2024PhDT}. The HEPAT model is developed based on the one-dimensional PATH model \citep{Zank+etal+2000,Li+2003,Li+2005ions} and two-dimensional iPATH model \citep{Hu+etal+2017}, which are physics-based models for simulating particle acceleration and transport in gradual SEP events. {The PATH and iPATH models have been proven to be successful in modelling some large SEP events and explaining the longitudinal dependence of SEP intensity \citep[e.g.,][]{Verkhoglyadova+2009,Verkhoglyadova+2010,Hu+etal+2018,Ding+2020,Li2021,ding2022A&A...668A..71D,Ding2022}.}
The HEPAT model consists of three major modules: (1) MHD Module: coupled with European heliospheric forecasting information asset (EUHFORIA; \citet{pomoell2018euhforia}), which is a comprehensive data-driven coronal and heliospheric model specifically designed for space weather forecasting. This module simulates data-driven solar wind and CME eruptions;  (2) Acceleration Module: using shock information derived from EUHFORIA, this module calculates time-dependent shock acceleration based on DSA mechanism. It accounts for particle injection, self-generated wave intensity near the shock, maximum particle energy, particle diffusion and escape from the shock; (3) Transport Module: this module solves the 3D focused transport equation \citep{Skilling1971ApJ} using the backward stochastic differential equation method to obtain time-intensity profiles at virtual observers. 

The general simulation flow proceeds as follows.  First, in the MHD module, CME eruptions are simulated using the Cone model \citep{odstrcil2004numerical}, which treats the CME as a hydrodynamic plasma cloud with enhanced density and speed. It is inserted into the background solar wind with a constant speed and angular width. Second, following the methodology described in \cite{ding2022A&A...668A..71D}, we first identify shock positions from the EUHFORIA simulation, then derive key shock parameters, including the shock speed, the shock compression ratio, and the shock obliquity. These parameters are essential inputs for capturing dynamic evolution of the shock acceleration.  Third, in the acceleration module,  the steady-state DSA solution is calculated based on the inputs of shock parameters. The accelerated particles, which convect with the shock and diffuse downstream of the shock, are tracked. The instantaneous escaped particle distribution functions at the shock front are recorded as the source for the particle transport. Finally, the time-intensity profiles at a desired observer can be obtained from the transport module to compare with measurements.

A detailed description of the HEPAT model is provided by \cite{Ding2024PhDT}. Here, we briefly discuss the critical parameters relevant to this work. Solving the continuous, time-dependent shock acceleration is computationally demanding. Therefore, shock parameters derived from EUHFORIA, with a time step of 1 hour, are passed to the acceleration module to compute the steady-state DSA solution along the shock surface. One key parameter in understanding IVD events is the maximum particle energy of particle acceleration at the shock.  In the steady-state solution of the DSA mechanism, we assume shock parameters do not vary significantly over the shock dynamic timescale \citep{Zank+etal+2000}, defined as $t_{\rm dyn} = \frac{R}{dR/dt}$, where $R$ is the radial distance of the shock from the Sun. The maximum particle momentum, $p_{\rm max}$, is therefore determined by balancing the acceleration time \citep{Drury+1983} with $t_{\rm dyn}$:
\begin{equation}
t_{\rm dyn} = \int_{p_{\rm inj}}^{p_{\rm max}} \frac{3s}{s-1} \frac{\kappa_{\rm up}}{U_{\rm up}^{2}} \frac{1}{p} dp,
\label{eq:dynamic time}
\end{equation}
where $p_{\rm inj}$ is the injection momentum, $s$ is the shock compression ratio, $\kappa_{\rm up}$ is the particle diffusion coefficient upstream of the shock, and $U_{\rm up}$ is the upstream flow speed in the shock frame. The acceleration time downstream of the shock is neglected under the assumption that the downstream diffusion coefficient is significantly smaller than upstream values \citep{axford1981acceleration}. With these assumptions,  the instantaneous $p_{\rm max}$ can be regarded as an approximation to the time-dependent DSA solution.

{We follow the approach in PATH/iPATH models to obtain $\kappa_{\rm up}$, given by  $\kappa_{\rm up}=\kappa_{\parallel} \cos^2 \theta_{\rm BN} + \kappa_{\perp} \sin^2 \theta_{BN}$, where $\theta_{\rm BN}$ is shock obliquity angle and $\kappa_{\parallel}$ and $\kappa_{\perp}$ represent the parallel and perpendicular diffusion coefficients, respectively \citep[e.g.,][]{Zank+etal+2000,Rice+2003,Li+2005ions,Li+2012,Hu+etal+2017}.   To determine $\kappa_{\parallel}$ upstream of the shock, it is necessary to evaluate the wave intensity. For a CME-driven shock, Alfv\'{e}n waves driven by protons streaming from the shock front which serves as the primary source of upstream turbulence that confines particles near the shock. The concept of scattering particles by Alfv\'{e}n waves near a shock was initially proposed by \citet{Bell1978MNRAS.182..147B} and further developed into a coupled wave-particle quasi-linear model by \citet{Lee+1983}. In our work, we employ a steady state solution of the wave intensity in front of the shock given in  \citet{Gordon+1999}.  This approach enables us to derive both the wave action and the energetic particle spectrum in a time-dependent manner. The amplified waves are proportional to the flux of streaming accelerated protons, which itself is related to the injection speed $V_{\rm inj}$.   For particles to participate in the DSA process, their speeds must exceed an injection threshold so that they can scatter diffusively across the shock, which refers to the injection speed. A key challenge in DSA is determining how particles are injected from the thermal or superthermal plasma at the shock.
A classic approach, introduced by \citet{giacalone+1999} and \citet{Zank2006}, ensures small particle anisotropy when applying the Parker transport equation. In this method, the injection speed is determined by constraining the total anisotropy 
$\xi$  to be less than $1$. However, this approach requires knowledge of the diffusion coefficients, while the injection speed itself is needed to compute the amplified wave intensity at the shock front, creating a dependency issue. To circumvent this, \citet{Li+2012} proposed an analytical expression for the injection speed based on shock geometry and compression ratio, which is adopted in this work.
We then obtain $\kappa_{\perp}$ from $\kappa_{\parallel}$ using an analytical result derived from the extended nonlinear guiding center (NLGC) theory \citep{shalchi2010analytic}, which builds upon the original NLGC theory \citep{Matthaeus2003ApJL}. Detailed calculations of the injection speed, diffusion coefficients and the corresponding particle distribution function follow the methodology outlined in \cite{Hu+etal+2017}.  }

{Upstream of the shock, the particle intensity falls off exponentially with a momentum‐dependent diffusion length scale, $\lambda_{\rm diff}$, determined by the upstream diffusion coefficient \citep{Drury+1983}. Beyond a certain distance ahead of the shock, particles are assumed to escape and propagate along the interplanetary magnetic field. The escape length scale, $\lambda_{\rm esc}$, is typically taken to be $2-4$ times larger than $\lambda_{\rm diff}$, which itself depends on the turbulence strength \citep{Zank+etal+2000,Rice+2003,Li+2005ions}. This momentum‐dependent escape length naturally leads to different escape processes, effectively trapping low‐energy particles close to the shock due to their small diffusion length scale  \citep{Zank2006,ding2024modelling}. However, trapping high‐energy particles is more challenging because the corresponding wave intensity decays rapidly as the shock propagates outward. Although the momentum‐dependent escape process affects the spectrum of escaped particles, it does not alter the IVD feature, which requires the delayed release of high‐energy particles.
}

{
When particles escape from the shock, they propagate in the solar wind. The transport of escaping particles is described by the 3D focused transport equation.  In the quasi-linear theory (QLT) \citep{Jokipii+1966}, the pitch angle diffusion coefficient $D_{\mu \mu}$ is given by,
\begin{equation}\label{eq:dmiumiu}
D_{\mu \mu } =  \frac{2\pi^{2}\Omega^{2}(1- \mu^2)}{B^{2}v\mu}g^{\rm slab} \left ( k_{\parallel} \right ),
\end{equation}
where $\mu$ is pitch angle cosine, $B$ is magnetic field strength, $v$ is particle velocity, $g^{\rm slab} $ is the turbulence power spectrum in the solar wind,  $\Omega = \frac{e B}{\gamma  m}$ is proton gyrofrequency, and resonant wave number is $k_{\parallel} = \Omega (v|\mu|)^{-1}$. $g^{\rm slab}$ we used here is given by \cite{shalchi+2009+book},
\begin{equation}\label{eq:gslab}
g^{slab}(k_{\parallel})=\frac{C(\nu )}{2\pi} l_{\rm slab}\delta B^{2}_{\rm slab} \left ( 1 + k^{2} l^{2}_{\rm slab} \right )^{-\nu},
\end{equation}
where $l_{\rm slab}=l_{\rm c,slab}/(2\pi C(\nu))$ is the slab bendover scale, where correlation length $l_{\rm c,slab} = 1 \times 10^{9}\;$m. $\delta B^{2}_{\rm slab}$ is the strength of the slab magnetic field and the inertial range spectral index $s=2\nu=5/3$. The normalization factor $C(\nu )$ equals
\begin{equation}\label{eq:cnu}
C(\nu) = \frac{1}{2\sqrt{\pi}}\frac{\Gamma(\nu)}{\Gamma(\nu-1/2)},
\end{equation}
where $\Gamma(x)$ is the Gamma function. The parallel diffusion coefficient $\kappa_{\parallel}$ is commonly expressed as:
\begin{equation}\label{eq:kappa_para}
\kappa_{\parallel} = \frac{v^2}{8}\int^{+1}_{-1}\frac{(1-\mu^2)^2}{D_{\mu\mu}}d\mu.
\end{equation}
The perpendicular diffusion coefficient $\kappa_{\perp}$ is then derived from $\kappa_{\parallel}$ using the NLGC Theory \citep{Matthaeus2003ApJL,shalchi2010analytic},
\begin{equation}\label{eq:kappa_perp}
\kappa_{\perp} = \left [ \frac{\sqrt{3}}{3} va^{2}\pi C(\nu) \frac{\delta B^{2}_{2D}}{B^2_{0}} l_{2D}\right ]^{2/3}\kappa_{\parallel}^{1/3},
\end{equation}
where the square of the turbulence magnetic field follows a radial dependence of $\delta B^{2}\sim r^{\gamma}$ with  $\gamma=-3.5$.  The 2D bendover scale is typically assumed to be $l_{\rm 2D} = 0.1 l_{\rm slab}$.  We assume the ambient turbulence level $ \delta B^{2}/B^{2} $ to be $0.1$ at 1 AU with an $80 : 20$ ratio of two-dimensional to slab component energies. This gives a reference value of $\kappa_{\perp}/\kappa_{\parallel }=0.0017$ for $1$ MeV proton at $1\;$au. We note that the discussion regarding the influence of perpendicular diffusion in IVD events pertains solely to the transport process rather than to shock acceleration. The role of perpendicular diffusion at CME-driven shocks has been extensively discussed in \cite{Zank2006}.
}

\section{Results}\label{sec:results}
\subsection{Solar Orbiter Observations}\label{subsec:solo observation}

\begin{figure*}
\includegraphics[width=\textwidth]{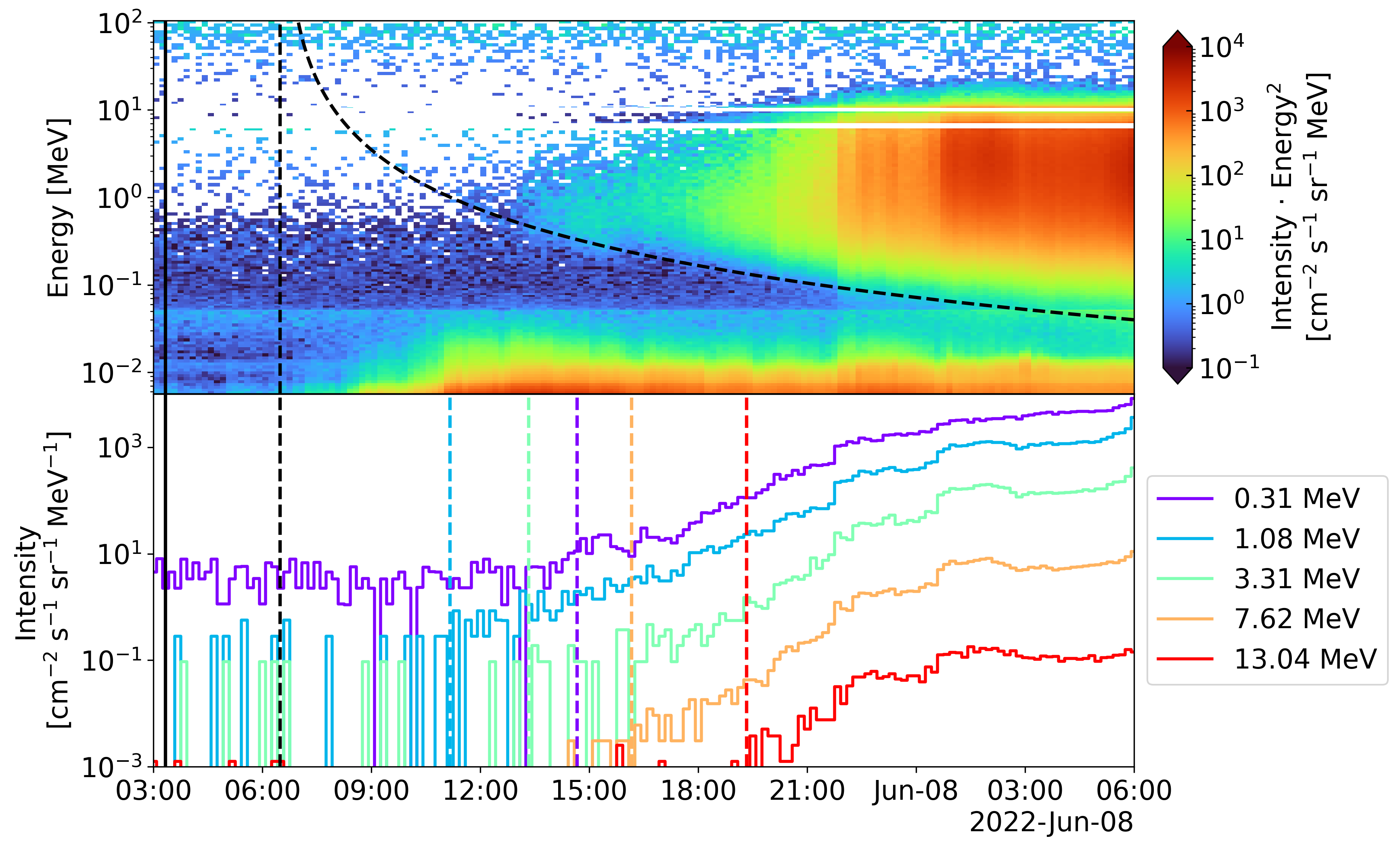}
\caption{Proton dynamic spectrum (upper panel) and time-intensity profiles (lower panel) for the  2022 June 7  event observed by STEP, EPT, and HET onboard SolO. The curved dashed line in the upper panel represents the onset times fitted using velocity dispersion analysis. The colorful vertical lines in the lower panel indicate the onset times for each energy channel. The vertical solid line marks the CME eruption time, while the vertical black dashed line denotes the release time derived from VDA.}
\label{fig1:time-intensity profile}

\end{figure*}
 
\begin{figure*}
\includegraphics[width=12.9cm]{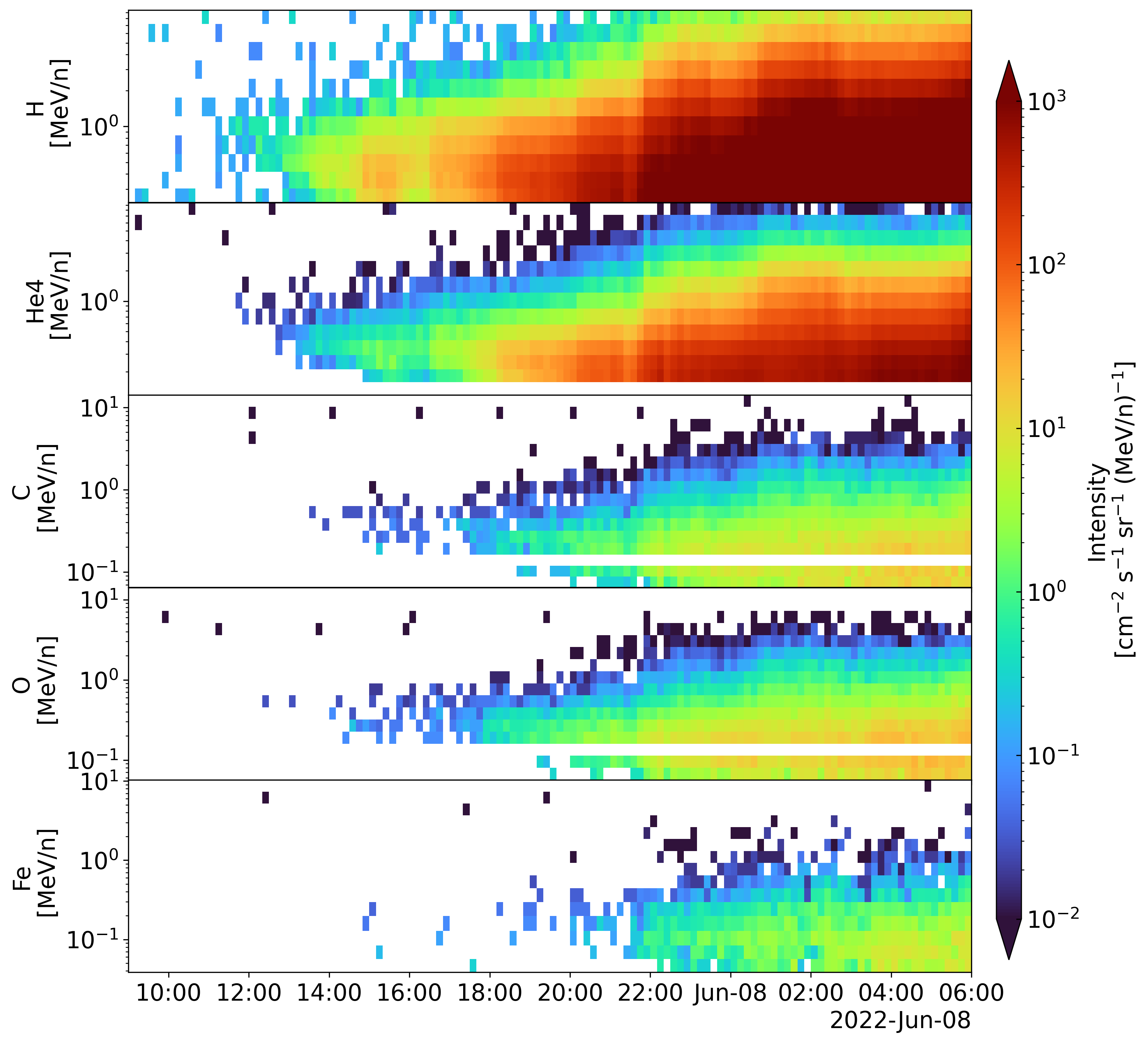}
\caption{Ion dynamic spectra for the 2022 June 7 events observed by SolO/SIS. From top to bottom, the panels show measurements of protons, helium-4, carbon, oxygen, and iron ions. }
\label{fig2:SIS spectra}
\end{figure*}

Figure~\ref{fig1:time-intensity profile} presents the 10-minute-averaged proton dynamic spectra and time-intensity profiles observed by SolO for the 2022 June 7  event.  
{As noted in Sect.~\ref{sec:methods}, the STEP and EPT instruments do not distinguish between ion species, and the measured flux is predominantly composed of protons. Therefore, we refer to Fig.~\ref{fig1:time-intensity profile} as proton dynamic spectra to differentiate them from the heavy-ion measurements provided by SIS (see Fig.~\ref{fig2:SIS spectra}).} 
The upper panel combines measurements from STEP with the averaged intensity obtained from four telescopes in EPT and HET. The color-coded bins on the grid represent $I \cdot E^2$, where $I$ is the particle intensity and $E$ is the mean energy of the corresponding channels. This representation flattens the energy spectrum, making it easier to visually identify velocity dispersion and inverse velocity dispersion in the SEP event.  The upper panel clearly reveals velocity dispersion in energy channels below $\sim 1\;\mathrm{MeV}$ and inverse velocity dispersion from approximately $ 1\;\mathrm{MeV}$ to $20\;\mathrm{MeV}$. The duration of the IVD spans roughly 10 hours, from 11:00 to 22:00 UT.  The transition energy from VD to IVD, referred to as the "nose energy" in this study, corresponds to the earliest onset time among the energy channels and is estimated to be around $1.1\;\mathrm{MeV}$. The nose energy can be regarded as the maximum particle energy at the acceleration site when the observer initially establishes magnetic connectivity to that site.  
To determine the particle release time using VDA method, we employ the Poisson-CUSUM method \citep{lucas1985counted,xu2020first} to find the onset time, focusing on energy channels between $0.1\;\mathrm{MeV}$ and $1\;\mathrm{MeV}$. The fitting results yield a release time of 2022-06-07 06:29 UT $\pm$ 23 minutes and a path length of $1.56 \pm 0.06\;\mathrm{au}$, is consistent with significant scattering in the interplanetary medium. This release time is approximately three hours later than the CME eruption time of 2022-06-07 03:36 UT, as determined from the first appearance time in LASCO/C2 field of view \footnote{\url{https://cdaw.gsfc.nasa.gov/CME_list/}}. The late release time indicates that SolO was not magnetically connected to the shock at the beginning of the CME eruption. Thus SolO may start the connection to the shock flank later when the CME reaches a certain height in the corona \citep{Rouillard2011ApJ,Rouillard2012ApJ}.

The bottom panel shows the time-intensity profiles of five energy channels along with their corresponding onset times. All energy channels exhibit a gradual intensity increase, reaching a plateau around 2022-06-07 22:00 UT. This behavior is characteristic of a typical eastern SEP event as seen from SolO, where the magnetic footpoint of SolO is located westward of the flare site.  The earliest onset time is approximately 2022-06-07 11:10 UT for the $1.08\;\mathrm{MeV}$ channel, with onset times gradually increasing for higher energies, demonstrating a clear inverse velocity dispersion. In summary, this event features a long-duration IVD with a nose energy around $1\;\mathrm{MeV}$.  Notably, this may be the clearest long-duration IVD event recorded by SolO before September 2024,  without perturbations of pre-events.  Thus, we use this clear example to explore the possible reasons behind the occurrence of IVD.

Figure~\ref{fig2:SIS spectra} presents 10-minute-averaged ion dynamic spectra for the  2022 June 7 event observed by SolO/SIS, for protons, helium-4, carbon, oxygen, and iron ions. First, all species exhibit a clear IVD with a long-duration of approximately 10 hours.  Second, the nose energy of these species shows a dependence on the charge-to-mass ($Q/M$) ratio, with the nose energy decreasing as the $Q/M$ ratio decreases, as evident from visual inspection. We note that the ion energy mentioned below is given per nucleon. Typically, the break energy in integrated ion spectra is used to deduce $Q/M$-dependence \citep{Desai2016ions}. However, such spectra represent a combination of continuous particle injections and are naturally dominated by the later phases of the SEP event, which tend to overshadow the properties of the early-stage shock. In contrast, the nose energy provides a probe of the initial properties of the shock. 
{ We note that the acceleration of heavy ions at the shock may be different compared to the proton acceleration. As suggested by \citet{Li+2005ions} and \citet{Verkhoglyadova2015PhR},  the self-generated waves at quasi-parallel and oblique shock are mainly excited by streaming protons while heavy ions do not significantly contribute to the excitation of fluctuations because of the relatively small number density. The maximum energy is also computed based on a resonance condition but only up to the minimum wave number excited by the energetic streaming protons.  For ions at a highly perpendicular shock, the maximum energy is independent of the resonance condition and depends mainly on the shock parameters and the upstream turbulence levels. This implies a $(Q/M)^2$ dependence of maximum particle energy of heavy ions at quasi-parallel shocks and a $(Q/M)^{1/2}$ or $(Q/M)^{4/3}$ dependence at highly perpendicular shocks \citep{Zank2006}.  }
If the nose energy represents the maximum particle energy accelerated at the shock shortly after the eruption, the $Q/M$ dependence of the nose energy may reflect the influence of the shock geometry on particle acceleration \citep{Li2009ApJ...702..998L}. However, this dependence can be further complicated by diffusive transport effects, which can lead to the earlier arrival of ions with smaller $Q/M$ ratios \citep{Mason+2012}. 
Determining the precise nose energy of heavy ions in this event is challenging due to low statistics for heavier ions (e.g., Fe). We do not explore this relationship in this study, but focus on investigating the possible reason of the formation of IVD.

\subsection{Model results}\label{subsec:model}
\begin{figure*}
\includegraphics[width=\textwidth]{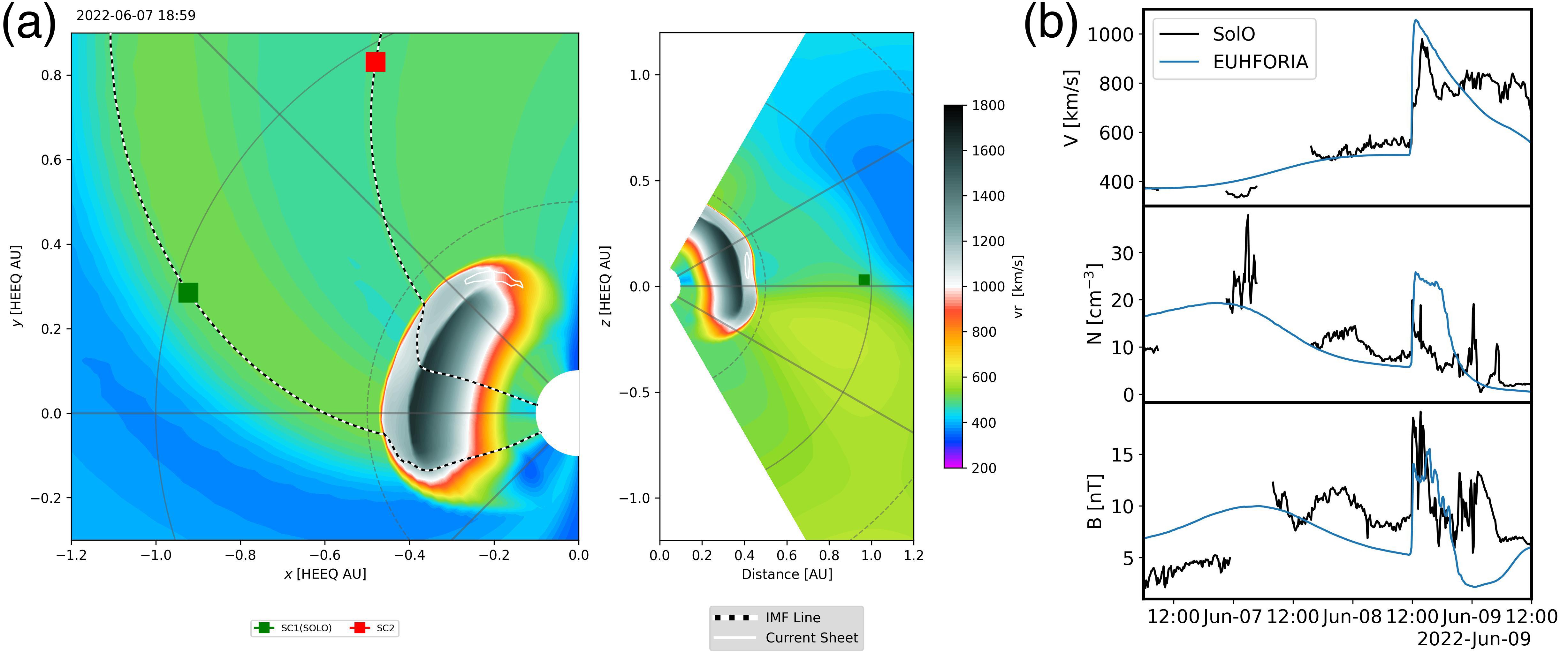}
\caption{  Panel (a): Equatorial (left) and meridional (right) snapshots of the radial solar wind speed from EUHFORIA. Panel (b): Comparison of in-situ plasma and magnetic field between observation and EUHFORIA simulation for the 2022 June 7 event. The panels present, from top to bottom, the solar wind proton number density, the solar wind speed, the magnetic field magnitude. The blue lines and the black lines are the EUHFORIA simulation results and measurements from SolO, respectively. }
\label{fig8:euhforia snapshot}
\end{figure*}

Figure~\ref{fig8:euhforia snapshot} (a) shows snapshots of the radial speed of the solar wind between $0.1\;$au to $2\;$au from EUHFORIA simulations for this event, presented in the Heliocentric Earth Equatorial (HEEQ) coordinate system. 
Since this eruption is a backside event from the Earth, accurate measurements of the CME parameters are not available. To estimate the eruption direction of the CME, we approximate its direction based on the flare location observed by SolO/Extreme Ultraviolet Imager (EUI; \citealt{rochus2020solar}). The CME speed and density in the Cone CME model are fine-tuned to match the shock arrival time observed at SolO. The parameters of the Cone CME used in the simulation are listed in Table~\ref{table1-cone}.  We acknowledge that the CME parameters are not tightly constrained by observations. However, the primary objective of the simulation in this study is to qualitatively investigate the mechanisms responsible for the observed IVD event, rather than to achieve precise quantitative modeling of the CME dynamics.  
In the snapshots, the propagation speed at the western portion of the CME is faster than that of the eastern portion. Such an asymmetric expansion of CME is due to variations in upstream solar wind conditions, that is, the CME expands faster in fast streams. This is also evident from the meridional slices, where a faster expansion of the CME towards the southern hemisphere is observed though the eruption direction of the CME is around 20 degrees in latitude. We note that the expansion of the CME and its driven shock plays an important role in particle acceleration and the resulting IVD (see more discussion below).   In the equatorial plane snapshots, two observers are denoted: one is SolO (SC1, shown in green), located at 163$^{\circ}$ in longitude and at 0.96 au initially connected to the western flank of the shock, and the other is the virtual observer (SC2, red), located 115$^{\circ}$ in longitude and also at 0.96 au, initially connected to the shock nose. In this study, we compare the model results between these two spacecraft to explore the possible reasons behind the observed inverse velocity dispersion.

  We further examine the plasma and magnetic field measurements by SolO.  The corresponding simulated time series of solar wind parameters are shown in Fig.~\ref{fig8:euhforia snapshot} (b).  From the comparison between modelled and observed time series,  SolO likely passed through a stream interaction region and entered a fast solar wind stream before the flare eruption, which is evident from the number density peak observed around June 7, 02:00 UT. Before the shock arrival, the plasma and magnetic field conditions are relatively undisturbed in the observation. Following the shock passage, a distinct sheath structure and a magnetic cloud are observed. The simulated shock arrival time and the magnitudes of the number density and flow speed near the shock are consistent with the observations, providing confidence that EUHFORIA simulations reasonably capture the shock propagation. However, we note that the Cone model used in the simulations does not include the magnetic flux rope.  Consequently, the model is unable to accurately reproduce the observed magnetic field disturbances, resulting in simulated magnetic field magnitudes during the passage of the CME that are not directly comparable to those recorded by SolO.

\begin{table}
\caption{\label{table1-cone}Input parameters of the Cone CME model in the EUHFORIA}
\centering
\begin{tabular}{lccc}
\hline\hline
Parameter &Value\\
\hline
Insertion time           &  2022-06-07T06:00:00  \\
Insertion latitude (HEEQ)      & 18$^\circ$  \\
Insertion longitude (HEEQ) &  168$^\circ$ \\
Half-width        &  55$^\circ$\\
Speed           &  1600 km/s \\
Density     &   $1.4$ $\times$ $10^{-18}$ kg m$^{-3}$\\
Temperature      &  $2.0$ $\times$ $10^6$ K\\

\hline
\end{tabular}
\end{table}

\begin{figure*}
\includegraphics[width=\textwidth]{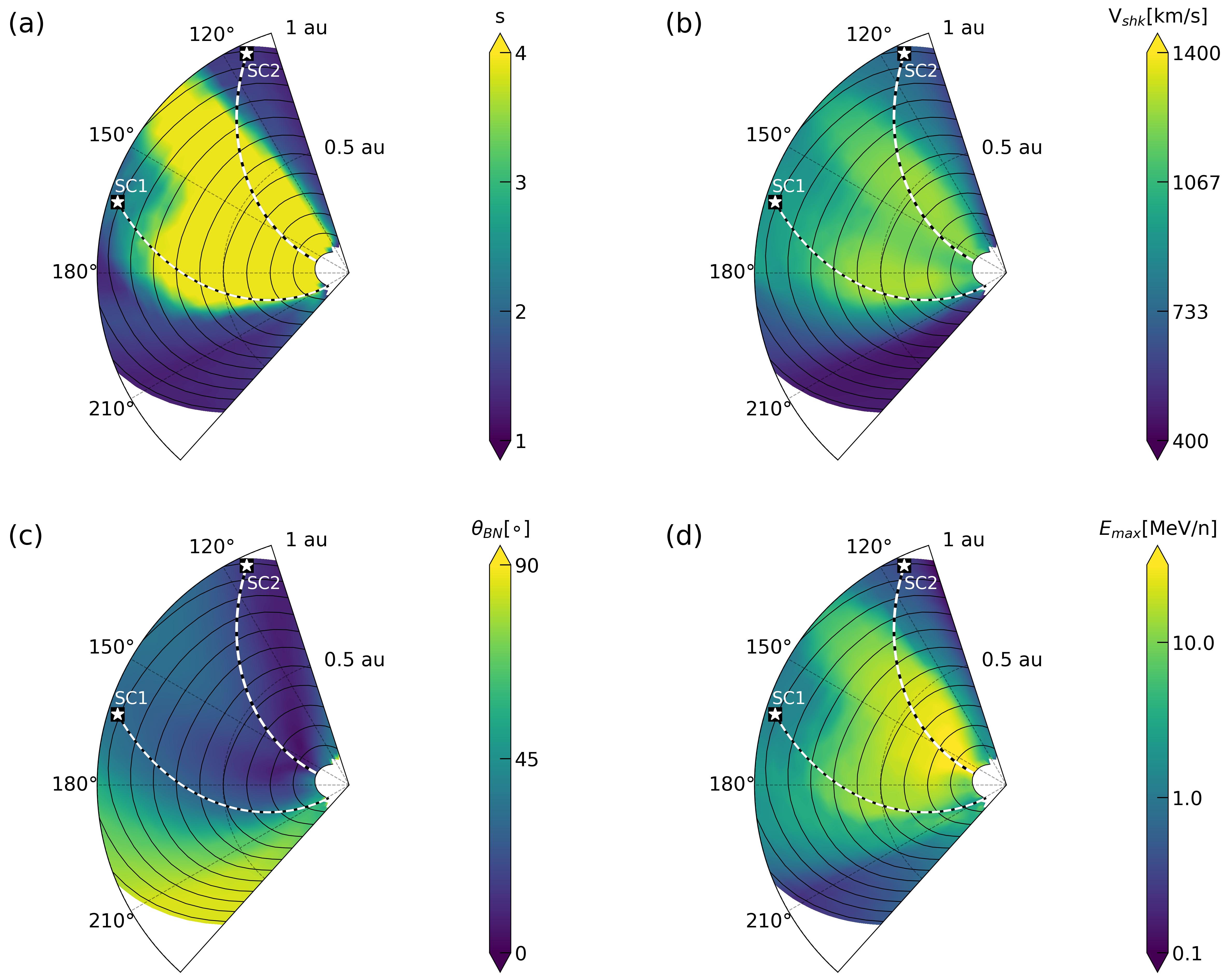}
\caption{The evolution of shock location and shock parameters (shock compression ratio (a), shock speed (b), shock obliquity angle (c) and maximum proton energy (d)) in the equatorial plane. The black solid curves show the shock front at different time steps. The color schemes are for different shock parameters along the shock front. The white dashed curves represent the Parker field lines passing through SC1 (SolO) and SC2. }
\label{fig9:shock paramters}
\end{figure*}

To illustrate the evolution of the shock profile, Figure~\ref{fig9:shock paramters} shows the time history of key shock parameters in the equatorial plane. The four panels represent the density compression ratio ($s$), the shock speed ($V_{\rm shk}$), the shock obliquity ($\theta_{\rm BN}$), and the maximum proton energy ($E_{\rm max}$). For this analysis, it is assumed that SolO is located in the solar equatorial plane, given its small latitude of $1.8^{\circ}$. The upper left panel depicts the evolution of the shock compression ratio. The highest compression ratios are observed at the shock nose, while the values decrease toward the shock flanks.  SolO initially connects to the shock flank, near the edge of the strong shock while SC2 first connects to the shock nose, where the compression ratio remains consistently high. 
The upper right panel shows the evolution of shock speed. Two distinct regions of high shock speed are evident. One region corresponds to shock expansion toward the eastern flank, while another exhibits significant acceleration with time near the longitude of $180^{\circ}$.  Near the Sun, the shock flank may be underdeveloped and can undergo further expansion as the shock propagates outward. This process is driven by the interaction between the expanding shock and the ambient solar wind, which enhances the lateral expansion of the shock structure. As a result,  SolO initially connects to the shock flank, where the shock speed is lower, and subsequently transitions to the shock nose, where the shock speed is higher.  The bottom left panel illustrates the evolution of the shock obliquity angle  $\theta_{\rm BN}$. A smooth transition in shock geometry is observed as the  $\theta_{\rm BN}$ increases from the eastern shock flank (quasi-parallel) to the western flank (quasi-perpendicular). 
The bottom right panel shows the maximum proton energy ($E_{\rm max}$) along the shock front. This panel highlights how SolO initially connects to the shock flank, where $E_{\rm max}$ is lower, before gradually connecting to regions with higher $E_{\rm max}$. In contrast, SC2 connects initially to the shock nose, where higher $E_{\rm max}$ are present immediately following the eruption. Overall, this figure clearly illustrates the evolving nature of shock parameters and their role in shaping the particle acceleration environment, demonstrating the importance of magnetic connectivity in determining the observed SEP characteristics.

\begin{figure*}
\includegraphics[width=\textwidth]{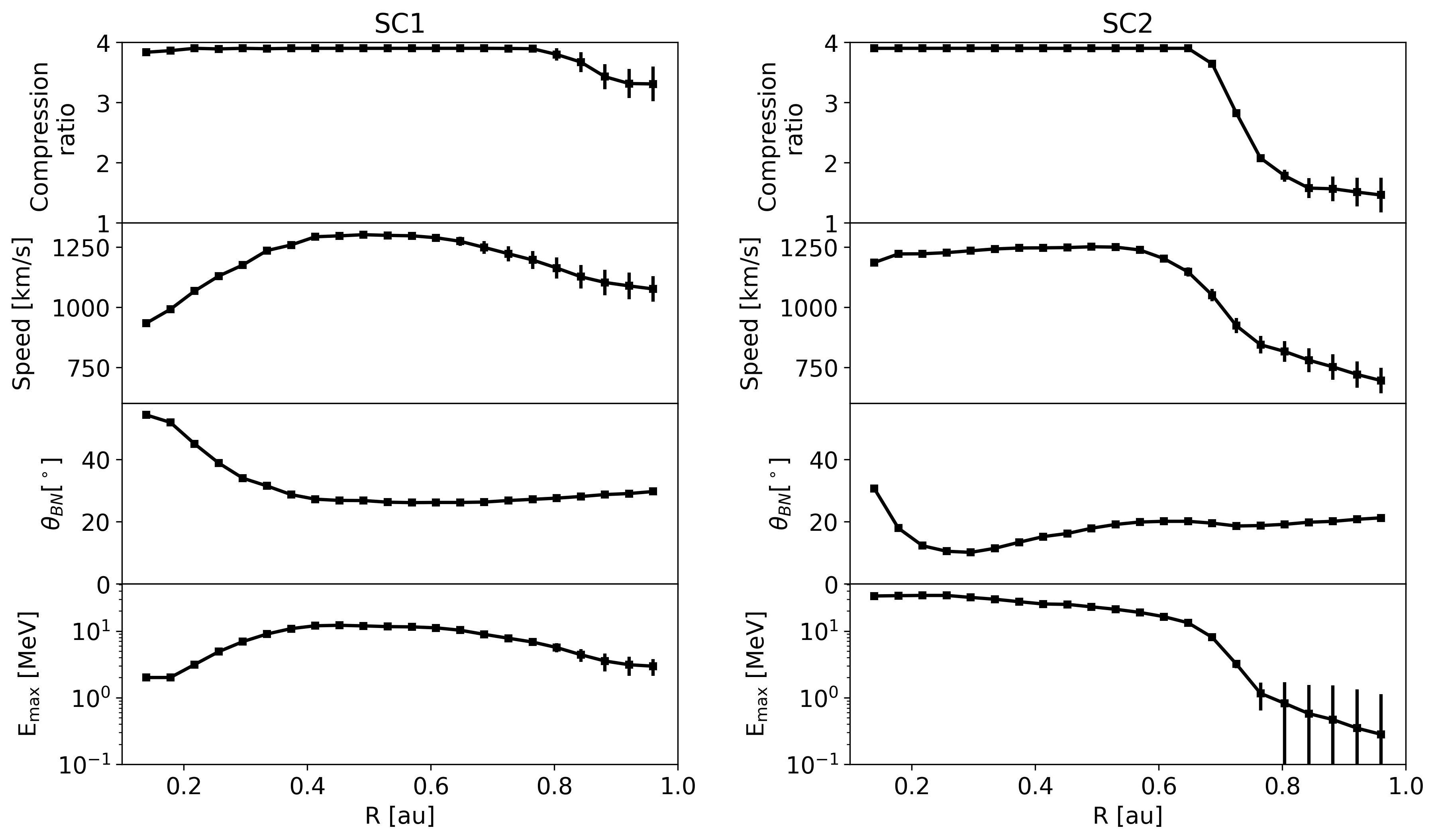}
\caption{Radial evolution of shock parameters along the field line connecting to SC1(SolO) and SC2. From top to bottom, are shock compression ratio, shock speed, shock obliquity angle and maximum proton energy. The error bars represent the uncertainty in field line connection, corresponding to a $2^\circ$ angular deviation near the observer.  
}
\label{fig:shock_params_fl}
\end{figure*}

To clearly demonstrate the different shock connectivities, Figure~\ref{fig:shock_params_fl} compares the shock parameters along the magnetic field lines connecting to SC1 (SolO) and SC2. Both virtual observers initially connect to regions of the shock with high compression ratios, which gradually decrease over time. The most significant difference between the two observers is the evolution of the shock speed.  SC1 initially connects to the shock flank and shifts toward the shock nose as the shock propagates. The shock speed along SC1's path line gradually increases until $\sim 0.4\;\mathrm{au}$, after which it plateaus and gradually decreases until the shock arrival. In contrast, SC2 initially connects directly to the shock nose, where the shock speed remains nearly constant around  $1200\;$km/s up to $0.6\;\mathrm{au}$, thereafter SC2's path line connects to the shock flank farther away from the nose.  The increase in shock speed observed along SC1’s path line is attributed to the shift from shock flank to shock nose and faster shock expansion in the fast solar wind.  Regarding shock geometry, SC1 initially connects to an oblique shock with $\theta_{BN} \sim 50^{\circ}$ following the onset, and gradually transitions to a more quasi-parallel shock as the connection shifts. Conversely, SC2 remains connected to a quasi-parallel shock throughout the period before the shock arrival.  The variations in the shock parameters eventually influence the evolution of the maximum proton energy $E_{\rm max}$ observed at the two virtual spacecraft. SC1 initially connects to the shock flank, where $E_{\rm max} \sim 2\;\mathrm{MeV}$. As the connection moves closer to the shock nose, $E_{\rm max}$ peaks around $10\;\mathrm{MeV}$ before declining. In contrast, SC2's initial connection to the shock nose results in a higher initial $E_{\rm max} \sim 30\;\mathrm{MeV}$, which subsequently decreases as the connection shifts toward the shock flank. This is a result of the higher compression ratio and shock speed at the nose close to the Sun, providing more efficient particle acceleration early in the event.
Overall, these comparisons highlight the evolving maximum particle energy in different magnetic connections and underscore the importance of shock evolution in determining particle acceleration efficiencies at different locations of the shock.

This is a key result in explaining the delayed onset of high-energy particles, driven by the observer’s evolving magnetic connection to regions of the shock with increasing $E_{\rm max}$ as the shock propagates through the inner heliosphere. As the observer initially connects to the shock flank, where $E_{\rm max}$ is lower, only lower-energy particles are detected. Over time, as the magnetic connection moves to a region along the shock front where $E_{\rm max}$ is higher, particles with higher energies begin to arrive, resulting in the observed inverse velocity dispersion.

\begin{figure*}
\includegraphics[width=\textwidth]{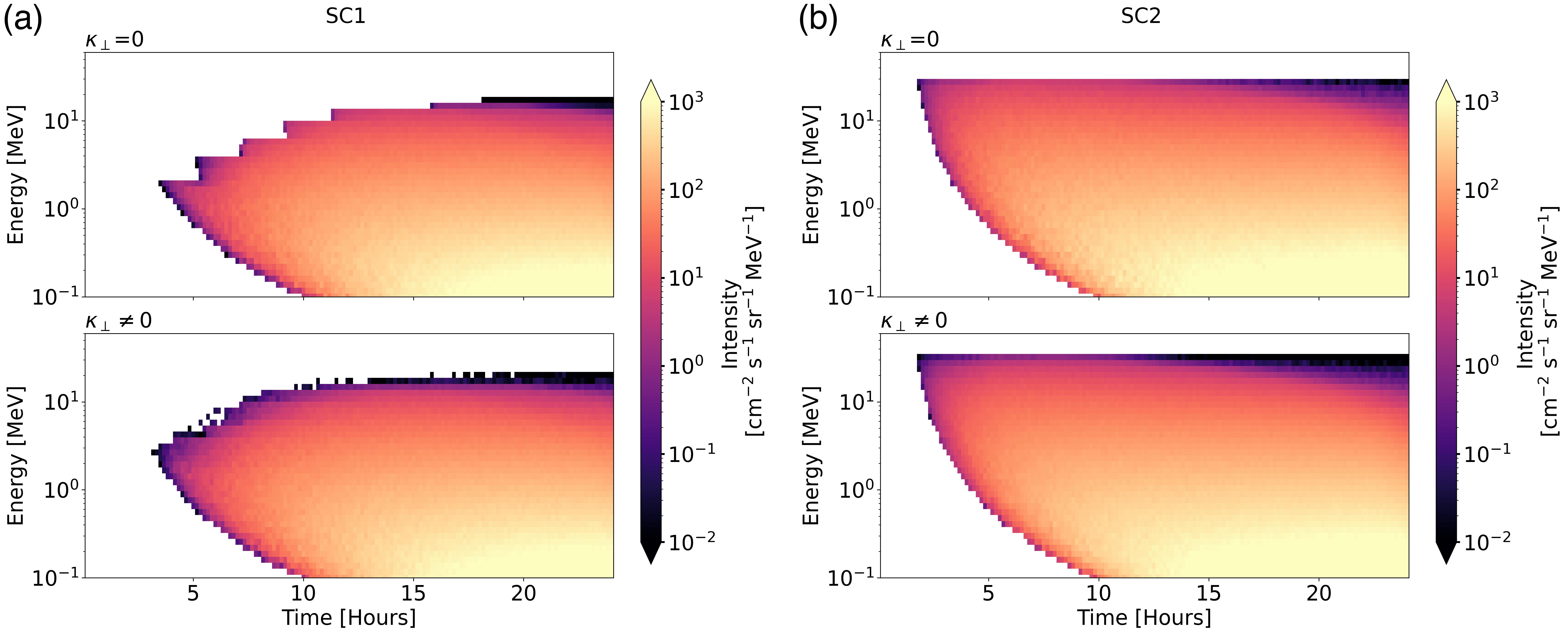}  
\caption{Modelled proton dynamic spectra observed at SC1 (a) and SC2 (b). Upper panels show the case without cross-field diffusion and lower panels show the case with cross-field diffusion. }
\label{fig:modelled dynamic spectra}
\end{figure*}

Figure~\ref{fig:modelled dynamic spectra} shows the modeled dynamic spectra as observed by SC1 and SC2 within 24 hours after the CME eruption. The upper panels represent the case without cross-field diffusion in the transport process, providing a straightforward interpretation of how shock connectivity drives the observed VD and IVD. 
In the upper left panel, the clear IVD pattern is evident, with a nose energy around $2\;\mathrm{MeV}$. Above this energy, the particle arrival time gradually increases with energy, exhibiting as inverse velocity dispersion. This behavior directly correlates with the evolution of the maximum proton energy along the magnetic field line connecting to SC1. Since cross-field diffusion is not included, accelerated particles can only propagate along the field line. As the maximum particle energy along this connection increases over time, higher-energy particles arrive later, explaining the delayed onset of the higher-energy particles. The step-like jumps shown in the IVD part are caused by the discrete time-dependent acceleration source as discussed in Sect.~\ref{sec:methods}. 
In contrast, the upper right panel, representing SC2, shows a distinct VD pattern. SC2 maintains a direct connection to the shock nose following the eruption, allowing for immediate access to the highest-energy particles. As a result, a typical velocity dispersion is observed, with higher-energy particles arriving first, followed by lower-energy particles.  

The lower panels display the modeled results which include cross-field diffusion in the transport process. We assume a reference value of $\kappa_{\perp}/\kappa_{\parallel} = 0.0017$ for 1 MeV protons at $1\;$au as discussed in Sect.~\ref{sec:methods}.  Comparing the SC1 results (left panels), it is evident that the nose energy is slightly higher when cross-field diffusion is included. This occurs because cross-field diffusion enables the observer to sample a broader region of the shock front, rather than being restricted to a single point along the magnetic field line. Consequently, higher-energy particles from regions of the shock with greater $E_{\rm max}$ can propagate to the observer, resulting in a higher nose energy. 
Moreover, cross-field diffusion reduces the duration of the IVD phase compared to the case without cross-field diffusion since higher-energy particles initially released from different regions of the shock can reach the observer earlier by crossing field lines. This highlights an important insight: to observe clear IVD signatures, the influence of cross-field diffusion must be limited. Large cross-field diffusion (e.g., $\kappa_{\perp}/\kappa_{\parallel} > 0.01$ ) may mask IVD by allowing rapid propagation of high-energy particles across magnetic field lines.   This finding aligns well with the results presented in \citet{Kouloumvakos2025}.  
In the case of SC2, where the observer is already well connected to the shock nose, the inclusion of cross-field diffusion does not change the results too much. As demonstrated by \citet{ding2022A&A...668A..71D}, cross-field diffusion plays a more critical role for poorly connected observers, while its effect on well-connected events remains relatively minor. Indeed, comparing dynamic spectra of 4 individual EPT telescopes, as shown in Appendix~\ref{appendix:EPT}, confirms that this event was anisotropic at Solar Orbiter. These results emphasize that IVD signatures are closely tied to the evolving shock connectivity along the magnetic field line.

\begin{figure*}
\includegraphics[width=\textwidth]{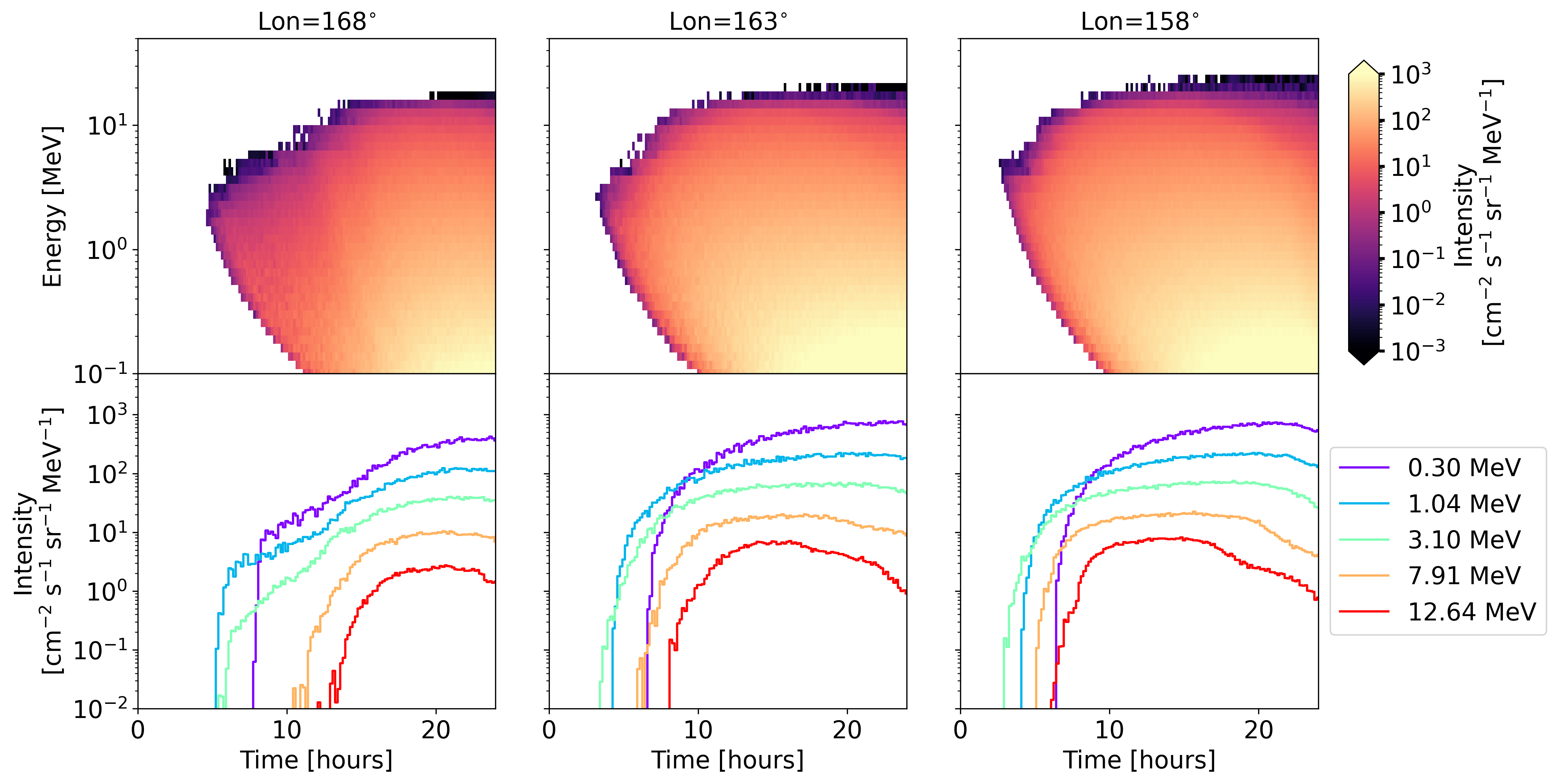}  
\caption{Modelled proton dynamic spectra and time-intensity profiles at longitudes of $168^{\circ}$, $163^{\circ}$ (SolO) and $158^{\circ}$. }
\label{fig:ivd-3c}
\end{figure*}

 To further investigate the uncertainty in magnetic connectivity, Figure~\ref{fig:ivd-3c} compares the dynamic spectra and time-intensity profiles at three longitudes, separated by 5 degrees around SolO’s position (Lon=163$^{\circ}$). Five energy channels are selected to match those in Fig.~\ref{fig1:time-intensity profile} for comparison.   These results include the effects of perpendicular diffusion, as discussed in Sect.~\ref{subsec:model}, to better reproduce the gradual enhancement of particle intensity observed in-situ. 
In the upper panels, variations in nose energy and the duration of the IVD are evident across different longitudes. As expected, the nose energy increases when the magnetic footpoint is closer to the shock nose and the duration of the IVD become shorter. The modelled time-intensity profiles at SolO in the middle panel show a more rapid enhancement of particle intensity compared to observations. 
A more consistent match with observations is found at a longitude of $168^{\circ}$, where the modelled results display a slower enhancement and a longer IVD duration. This discrepancy suggests that SolO may have poorer magnetic connectivity to the shock initially, with the limited cross-field diffusion contributing to the gradual increase in particle intensity during the early phase of the event.
These results indicate that IVD events may be more pronounced in poorly-connected SEP events.
We note that the modelled results exhibit a significantly harder energy spectrum compared to the observations, suggesting that the EUHFORIA simulation overestimated the shock compression ratio at the region magnetically connected to the observer. The current simulation is primarily constrained by the shock arrival time observed by SolO, making it challenging to achieve a lower shock compression ratio in the model.

\begin{figure*}
\includegraphics[width=\textwidth]{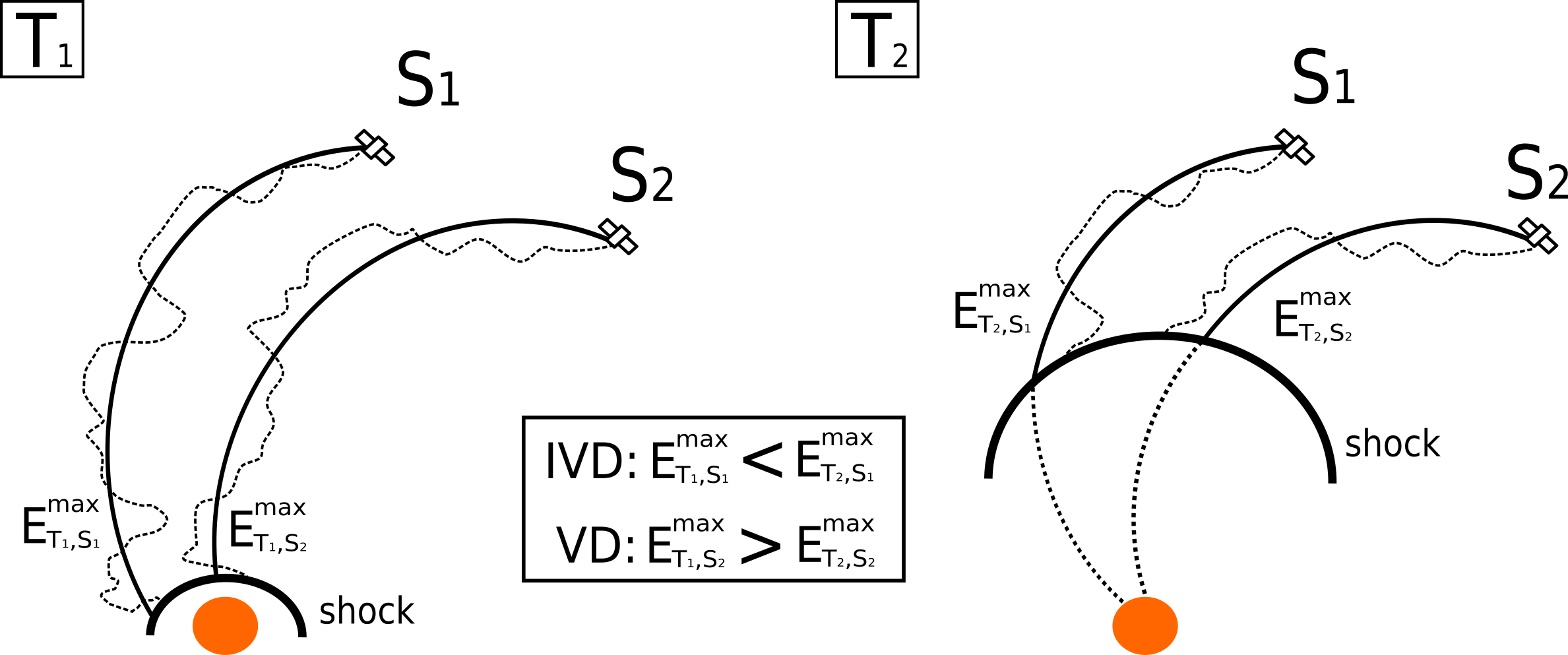}  
\caption{A schematic of shock connectivity at two time steps, $T_1$ and $T_2$. $E_{T_x,S_x}^{max}$ indicates the maximum proton energy connected to observer $S_x$ at time $T_x$. }
\label{fig:sketch}
\end{figure*}

To summarize the findings of this work, Figure~\ref{fig:sketch} shows a schematic representation of shock connectivity at two distinct time steps, $T_1$ and $T_2$. $T_1$ corresponds to the CME eruption time, while $T_2$ represents a later stage during the shock propagation. In both panels, two observers, $S_1$ and $S_2$, connect to the shock front through the mean magnetic field line (solid lines) and meandering field lines (dashed lines). $S_1$ and $S_2$ represent poorly connected and well-connected observers, respectively. The notation $E_{T_x,S_x}^{\rm max}$ indicates the maximum proton energy connected to observer $S_x$ at time $T_x$. For $S_1$, the magnetic connection at $T_2$ corresponds to a higher maximum particle energy as $ E_{T_2,S_1}^{\rm max} > E_{T_1,S_1}^{\rm max}$. In contrast, $S_2$ connects to the highest $E_{\rm max}$ at $T_1$, with $E_{T_1,S_2}^{\rm max} > E_{T_2,S_2}^{\rm max}$. As a result, observer $S_1$ is expected to detect inverse velocity dispersion at later times, while observer $S_2$ will observe velocity dispersion throughout the event.
It is important to note that $E_{\rm max}$ does not necessarily correspond to the maximum particle energy at the shock location directly connected to the observer. Instead, particles with higher energies may originate from different regions of the shock and propagate across magnetic field lines through cross-field diffusion. This highlights the critical role of limited cross-field diffusion in shaping the observed characteristics of IVD events.

\section{Conclusion and Discussion}\label{sec:conclu}

In this study, we investigated the inverse velocity dispersion in the SEP event observed by Solar Orbiter on 2022 June 7. This event exhibits a distinct and long-duration IVD signature, persisting for approximately 10 hours across the proton energy range from $1\;$MeV to $20\;$MeV. Measurements of ions also reveal a clear IVD signature, with varying nose energies corresponding to different ion species. To explore the possible mechanisms responsible for the formation of IVD, we employed the HEPAT model to simulate shock acceleration and particle transport for this SEP event. The simulation results suggest that the observed IVD is closely linked to the evolving maximum particle energy at the shock front along the magnetic connectivity, highlighting the importance of considering both time-dependent shock acceleration and transport processes in the interpretation of IVD events.
Our main findings are summarized as follows:
\begin{enumerate}
    \item The long-duration IVD event is driven by evolving shock connectivity during shock propagation. As SolO’s magnetic connection to the shock transitions from the flank to the nose, the observed maximum particle energy ($E_{\rm max}$) increases, leading to the delayed arrival of high-energy particles and the formation of the IVD signature.  
    \item The effect of cross-field diffusion is limited for IVD events. Moderate cross-field diffusion may shorten the duration of IVD and increase the observed nose energy, but large cross-field diffusion may mask the IVD by allowing high-energy particles to propagate rapidly across magnetic field lines. 

    \item Interplanetary shock expansion may enhance acceleration efficiency at larger solar distances, creating favorable conditions for sustained particle acceleration. This process contributes to the development of IVD features by accelerating higher-energy particles at later times. 
    
\end{enumerate}

In this simulation, the shock is initiated at $0.1\;\mathrm{au}$, which limits our ability to model coronal shocks directly. It is important to note that shock expansion also happens frequently in the corona, as demonstrated by \cite{Kouloumvakos2025}, who highlighted the critical influence of expanding coronal shocks in generating IVD events. For the IVD events of 2022 September 5 observed by PSP, \cite{Kouloumvakos2025} found that the coronal shock is initially sub-critical and transitions to a super-critical state, which allows for generating higher-energy particles at later times.  Meanwhile, \cite{Kouloumvakos2025} also suggests that cross-field diffusion appeared negligible in the PSP event. The primary reason for limiting cross-field diffusion effects in IVD events is to preserve properties of shock evolution along the field line. Therefore, the combined findings of \cite{Kouloumvakos2025} and this study suggest that shock evolutions and evolving magnetic connectivity, from the corona to the heliosphere, are key factors in producing IVD signatures by promoting efficient acceleration at later stages of shock evolution. The PSP event occurs close to the Sun, primarily linked to local shock evolution in the corona. However, based on our simulation results, which account for IP shock acceleration, we propose that long-duration IVD events are more commonly associated with SEP events where the observer’s magnetic footpoint is located west of the flare site.    
This configuration might facilitate a gradual transition of magnetic connectivity from the shock flank to the shock nose, increasing the likelihood of encountering stronger shock regions at later times. However, this is not absolute, as solar wind conditions and shock evolution can be highly dynamic \citep{Wijsen2023ApJ...950..172W,ding2024modelling}. 
{Connectivity can be far more complex due to the enhanced turbulence around the shock as suggested in \cite{Zank+etal+2000}. This complexity in turn can influence the effective magnetic connectivity. The escaped particles may be from a site different from that suggested by the simulated shock geometry and the interplanetary magnetic field.}

In addition to the combined effects of magnetic connectivity and shock evolution, another potential explanation for IVD events is the longer acceleration time required for higher-energy particles by the DSA process, resulting in their delayed release. In our model, we account for the acceleration time in DSA, as described by Eq.~\ref{eq:dynamic time}, to calculate the maximum particle energy at the shock front. Notably, the instantaneous $E_{\rm max}$ at a traveling shock does not necessarily increase with longer acceleration times. {This is because, as the shock propagates outward, the decay in magnetic field strength increases the diffusion coefficient near the shock, thereby reducing acceleration efficiency \citep{Zank+etal+2000,Rice+2003,Zank2006}.} The role of acceleration time might be more significant in scenarios where the diffusion coefficient and shock properties remain relatively constant, which is more likely during a short period when the shock is still close to the Sun. While we do not explicitly explore the effects of acceleration time on IVD events in this study, a more comprehensive investigation of acceleration time can be found in \cite{Li+2025}. 

Given the natural explanation of IVD events and that these are seen at all heliocentric distances covered by SolO \citep{Li+2025}, one is inclined to ask oneself why such events have not been reported previously. We have checked data from Solar Electron and Proton Telescope (SEPT; \cite{mueller-mellin-etal-2008}) onboard STEREO and have, indeed, found a number of such events. However, they are not as easily recognizable because SEPT has less energy resolution than SolO/EPT, which probably explains why they have not been reported previously. Although IVD detections are becoming more frequent, particularly through the high-resolution SEP measurements from SolO/EPD,   understanding why IVD events remain relatively uncommon in SEP observations is equally critical. This study highlights two key lessons in this regard.
First, IVD events require the late release of high-energy particles, which may need the transition of magnetic connectivity from the weak shock to the strong shock. Such conditions might depend on specific factors, such as the significant IP shock expansions \citep{manchester2017physical}, CME-CME interactions \citep{lugaz2017interaction}.  Second, large cross-field diffusion can obscure IVD features by enabling particle transport across magnetic field lines efficiently, thereby mixing contributions from a wide range of sources along the shock. The effects of cross-field diffusion can vary significantly from event to event due to the different turbulence levels in the solar wind and can accumulate over larger solar distances, potentially explaining the relative scarcity of IVD in SEP events.

In conclusion, this study suggests that long-duration IVD events are driven by evolving shock connectivity during shock propagation, where the observed maximum particle energy increases along the magnetic field line over time. Currently, the underlying mechanisms responsible for IVD events remain an open question. {While an explanation based on acceleration time may not necessarily require a specific longitudinal location along the shock front, it likely requires proximity to a sufficiently young shock near the Sun. In contrast, a connectivity-based explanation may depend on a particular magnetic connection, more likely linking to the western flank of the shock at the onset and moving to the nose part later on. } Continued observational and modeling efforts are crucial to fully understand the factors driving IVD events and to refine our interpretations. To date, Solar Orbiter has detected over ten candidate IVD events. Future statistical studies will be pursued by exploring the relationship between IVD characteristics and shock properties.

\begin{acknowledgements}
     We thank Gary Zank for his helpful comments. 
     Solar Orbiter is a mission of international cooperation between ESA and NASA, operated by ESA. This work was supported by the German Federal Ministry for Economic Affairs and Energy and the German Space Agency (Deutsches Zentrum für Luft- und Raumfahrt, e.V., (DLR)), grant number 50OT2002. The UAH team acknowledges the financial support by Project PID2023-150952OB-I00 funded by MICIU/AEI/10.13039/501100011033 and by FEDER, UE.  The Suprathermal Ion Spectrograph (SIS) is a European facility instrument funded by ESA under contract number SOL.ASTR.CON.00004. Solar Orbiter post-launch work at JHU/APL and the Southwest Research Institute is supported by NASA contract NNN06AA01C.
     This research was supported in part through high-performance computing resources available at the Kiel University Computing Centre. 
\end{acknowledgements}

\bibliographystyle{aa} 
\bibliography{aams} 

\appendix

\section{ Supplementary figure of the 2022 June 7 event}
\label{appendix:EPT}
\begin{figure}
    \centering
    \includegraphics[width=\textwidth]{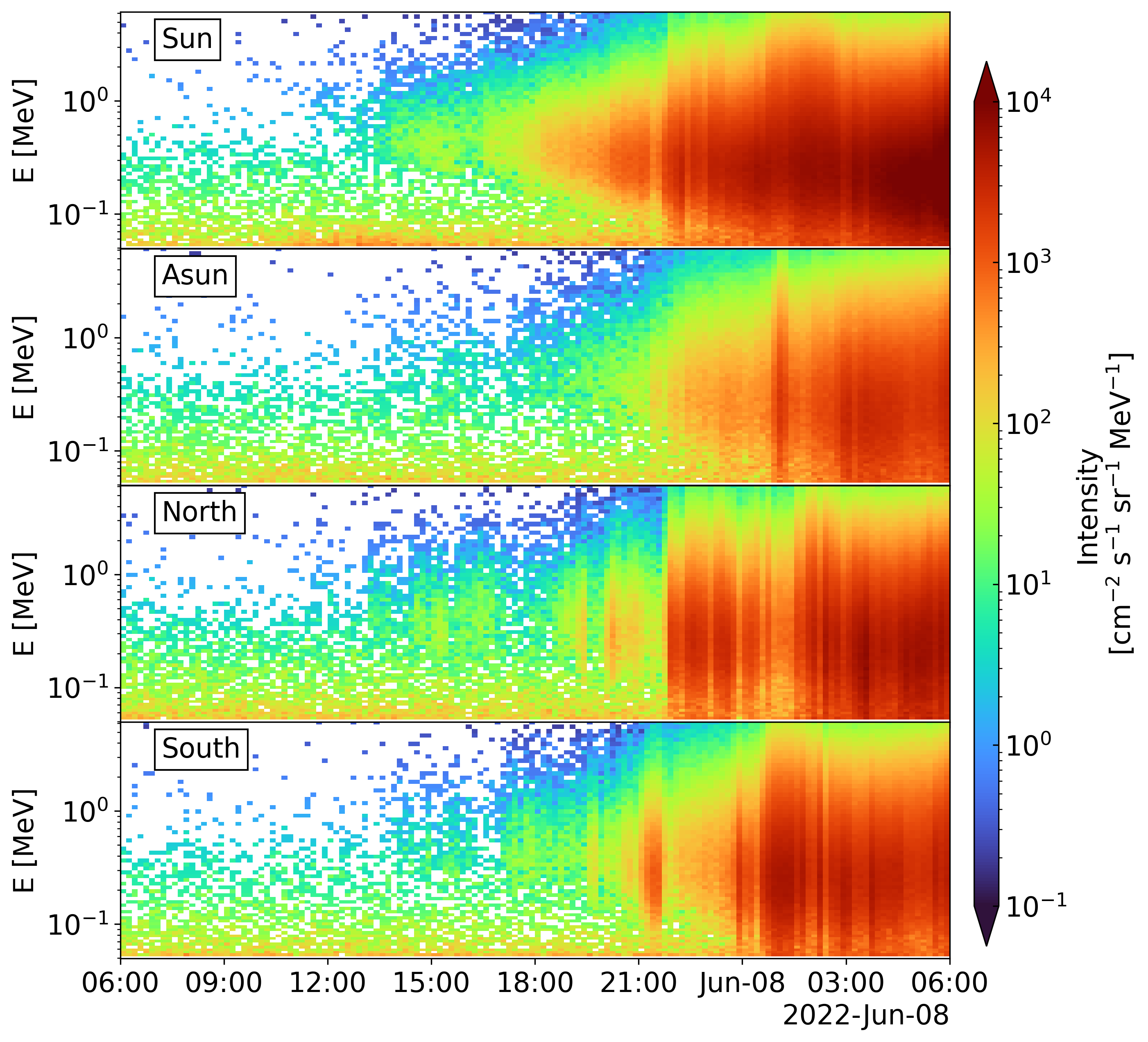}
    \caption{ Ion intensities for the  2022 June 7 event measured by the sun, anti-sun, north, and south telescopes of SolO/EPT.}
    \label{fig:EPT}
\end{figure}

Figure~\ref{fig:EPT} displays the color-coded ion intensities measured by the sunward, anti-sunward, north, and south telescopes of SolO/EPT. A comparison of the intensities across different telescopes reveals a pronounced anisotropy in the particle distribution, with the majority of particles streaming away from the Sun. This indicates a sustained and prolonged injection of energetic particles near the Sun, suggesting continuous particle acceleration and release over an extended period, associated with extended shock acceleration. 

\end{document}